\documentclass[12pt]{article}
\usepackage{graphicx}
\usepackage{multirow}
\usepackage{booktabs}
\usepackage{braket}
\usepackage{subfigure}
\usepackage{epstopdf}
\usepackage{float}
\usepackage{amsmath,amssymb,amstext,amsfonts,array,hhline,tabularx,graphicx,epstopdf,}
\usepackage[numbers,sort&compress]{natbib}
\makeatletter

\renewcommand{\maketag@@@}[1]{\hbox{\m@th\normalsize\normalfont#1}}%
\UseRawInputEncoding

\makeatother

\begin{document}
\parskip=3pt
\parindent=18pt
\baselineskip=20pt
\setcounter{page}{1}

\title{
Quantum Circuit Implementation of Two Matrix Product Operations and Elementary Column Transformations
\author{\ Yu-Hang Liu $^{1}$ , Yuan-Hong Tao $^{1}$ \footnote{Corresponding author: Yuan-Hong Tao E-mail: taoyuanhong12@126.com}, Jing-Run Lan $^{1}$ Shao-Ming Fei $^{2}$ \\
\footnotesize{1. School of Science, Zhejiang University of Science and Technology, Hangzhou 310023, China}\\
\footnotesize{2. School of Mathematical Sciences, Capital Normal University, Beijing 100048, China  }\\
}}
\date{}
\maketitle
\date{}

\textbf{\footnotesize{Abstract:}} \footnotesize{ This paper focuses on quantum algorithms for three key matrix operations: Hadamard (Schur) product, Kronecker (tensor) product, and elementary column transformations¡ªeach.
By designing specific unitary transformations and auxiliary quantum measurement, efficient quantum schemes with circuit diagrams are proposed. Their computational depths are: \(O(1)\) for Kronecker product; \(O(\max(m,n))\) for Hadamard product (linked to matrix dimensions); and \(O(m)\) for elementary column transformations of \(2^n \times 2^m\) matrices (dependent only on column count).
Notably, compared to traditional column transformation via matrix transposition and row transformations, this scheme reduces computation steps and quantum gate usage, lowering quantum computing energy costs.
 }

\textbf{{Keywords:}} {Quantum circuit, Unitary transformation, Kronecker product, Hadamard product, Elementary column transformation}\\

\begin{normalsize}

\section{Introduction}

\vspace{0.2cm}

Quantum algorithms have gained attention for their disruptive potential, leveraging qubit superposition and entanglement to achieve exponential speedup in solving complex problems. Landmark examples include Shor¡¯s factorization \cite{PW,PW1} (breaking classical cryptosystems), Deutsch¡¯s parallel algorithm \cite{DD} (proving quantum parallelism), Grover¡¯s search \cite{GLK} (boosting search efficiency), and quantum Fourier transform \cite{CD,WPF,NC} and phase estimation \cite{NC,WWL} (supporting quantum simulation and error correction), all highlighting quantum computing¡¯s unique value.

Matrix operations are core to quantum algorithm performance: quantum states are vectors, and quantum gates correspond to unitary matrix transformations. Phenomena like superposition, entanglement, and interference rely on matrix operations for accurate description, with quantum logic gates (e.g., Hadamard, CNOT) mapping to specific unitary matrices. This mechanism enables quantum computing¡¯s parallel processing and classical performance breakthrough.
The Hadamard (Schur) product, Kronecker (tensor) product, and elementary matrix transformations are foundational in cutting-edge fields: they support quantum gate design, signal filtering, and machine learning (e.g., attention mechanisms). Elementary transformations also underpin linear equation solving and matrix inversion.

In 2008, Duchamp et al. \cite{DGP} studied rational Hadamard products via Heisenberg-Weyl algebra.  
In 2021, Liming Z. et al. \cite{ZZR} developed QMAT for quantum linear algebra (supporting matrix addition, Kronecker/Hadamard products); Datt M. S. \cite{DMS} explored Kronecker products under quantum \(GL_2\).  
In 2022, Jonas T. et al. \cite{THZ} used Kronecker products for quantum three-body problem calculations.  
In 2023, Wenjie L. et al. \cite{LL} proposed secure two-party quantum Kronecker product (S2QSP) with polynomial complexity.  
In 2024, Reffk M. et al. \cite{MAH} developed QTPD for unitary matrix Kronecker decomposition; Wentao Q. et al. \cite{QZKW} built a quantum algorithm system for matrix operations (e.g., multiplication, inversion); Zenchuk A. I. et al. \cite{ZQKW,ZBQW} designed algorithms for inner products, determinant calculation, and linear equation solving.  
In 2025, Zenchuk A. I. et al. \cite{FZQW,ZQW,ZQW2,ZQW3} proposed algorithms for matrix conjugate transpose, quantum superposition states, and Hermitian conjugation; our team \cite{LTLF} developed quantum schemes for row operations, trace calculation, and transpose using Toffoli gates and auxiliary measurements.

Building on prior work, this paper explores quantum implementations of matrix Kronecker product (via C-SWAP and register merging), Hadamard product (via custom unitary operators and auxiliary marking), and elementary column transformations. Unlike indirect column transformations (via row operations + transposition, high energy cost), we design a dedicated, energy-efficient algorithm to optimize matrix operation applications in quantum computing.
\medskip

\vspace{0.2cm}
\section{Preliminaries}

This section introduces two important matrix operations: the Kronecker product (tensor product) and the Hadamard product (Schur product).  

Let matrices \( A=(a_{ij}) \in \mathbb{C}^{m \times n} \) and \( C \in \mathbb{C}^{p\times q} \); then the Kronecker product of \( A \) and \( C \) \cite{NC} is defined as the block matrix:  
\begin{equation}\label{eq:kronecker}
A \otimes C =
\begin{bmatrix}
a_{11}C & \cdots & a_{1n}C \\
\vdots & \ddots & \vdots \\
a_{m1}C & \cdots & a_{mn}C
\end{bmatrix} \in \mathbb{C}^{mp \times nq}.
\end{equation}  

The Kronecker product has wide applications in fields such as high-dimensional tensor representation and quantum composite system modeling \cite{HJ,LAJ}. It possesses the following basic properties:  
\begin{enumerate}
    \item Associativity: \( (A \otimes B) \otimes C = A \otimes (B \otimes C) \);  
    \item Distributivity: \( A \otimes (B + C) = A \otimes B + A \otimes C \) (dimension matching is required);  
    \item Mixed product property: If matrix multiplication is feasible, then \( (A \otimes B)(C \otimes D) = (AC) \otimes (BD) \);  
    \item Inverse and transpose: \( (A \otimes B)^{-1} = A^{-1} \otimes B^{-1} \) (if \( A \) and \( B \) are invertible), and \( (A \otimes B)^\top = A^\top \otimes B^\top \).  
\end{enumerate}  

Let matrix \( B=(b_{ij}) \in \mathbb{C}^{n \times m} \); then the **Hadamard product** of \( A \) and \( B \) \cite{RC} is defined as:  
\begin{equation}\label{eq:hadamard}
(A \circ B)_{ij} = a_{ij} b_{ij}, \quad 1 \leq i \leq n, 1 \leq j \leq m,
\end{equation}  
that is,  
\begin{equation}
A \circ B =
\begin{bmatrix}
a_{11}b_{11} & a_{12}b_{12} & \cdots & a_{1m}b_{1m} \\
\vdots & \vdots & \ddots & \vdots \\
a_{n1}b_{n1} & a_{n2}b_{n2} & \cdots & a_{nm}b_{nm}
\end{bmatrix} \in \mathbb{C}^{n \times m}.
\end{equation}  

The Hadamard product has wide applications in fields such as covariance matrix correction, image pixel-level operations, and adaptive optimization algorithms \cite{SGPH}. It satisfies the following properties:  
\begin{enumerate}
    \item Commutativity: \( A \circ B = B \circ A \);  
    \item Combination with diagonal matrices: If \( D \) is a diagonal matrix, then \( DA \circ B = D(A \circ B) \);  
    \item Schur product theorem: If \( A \) and \( B \) are positive definite matrices, then \( A \circ B \) is also positive definite.  
\end{enumerate}

\section{Quantum Algorithms for the Hadamard Product and Kronecker Product of Matrices}

This section introduces quantum algorithms for the Hadamard product and Kronecker product of matrices, respectively.

\subsection{Hadamard Product of Matrices}

Let \( A^{(1)} \) and \( A^{(2)} \) be \( N \times M \) matrices, where the elements of matrix \( A^{(i)} \) are given by \( A^{(i)} = (a^{(i)}_{s_i t_i}) \) for \( i = 1, 2 \) (\( N = 2^n \), \( M = 2^m \); \( n \) and \( m \) are positive integers). The quantum algorithm circuit diagram for their Hadamard product is shown in Figure 1.  

\begin{figure}[h]
\centerline{\includegraphics[width=0.8\textwidth]{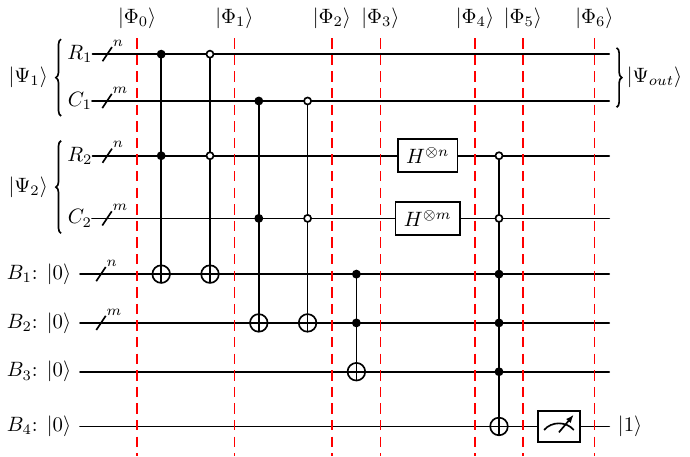}}
\caption{Quantum Circuit Diagram for the Hadamard Product}
\end{figure}

First, two \( n \)-qubit registers \( R_1 \) and \( R_2 \), as well as two \( m \)-qubit registers \( C_1 \) and \( C_2 \), are introduced. Among them, \( R_1 \) and \( C_1 \) are used to index the rows and columns of matrix \( A^{(1)} \), while \( R_2 \) and \( C_2 \) are used to index the rows and columns of matrix \( A^{(2)} \).  

The elements of the two matrices are encoded separately to obtain the following two pure states:  
\begin{eqnarray}
|\Psi_1\rangle &=& \sum_{s_1=0}^{N-1}\sum_{t_1=0}^{M-1}a^{(1)}_{s_1t_1}|s_1\rangle_{R_1}|t_1\rangle_{C_1},\\\nonumber
|\Psi_2\rangle &=& \sum_{s_2=0}^{N-1}\sum_{t_2=0}^{M-1}a^{(2)}_{s_2t_2}|s_2\rangle_{R_2}|t_2\rangle_{C_2},
\end{eqnarray}  
where \( |s\rangle \) denotes the binary representation of \( s \), and the coefficients satisfy the normalization condition \( \sum_{s_i t_i}|a^{(i)}_{s_i t_i}|^2 = 1 \) for \( i = 1, 2 \). The initial state of the entire system is constructed by combining the above two pure states via tensor product:  
\begin{eqnarray}
|\Phi_0\rangle &=&\sum_{s_1,s_2=0}^{N-1}\sum_{t_1,t_2=0}^{M-1} a^{(1)}_{s_1t_1}a^{(2)}_{s_2t_2}|s_1\rangle_{R_1}|t_1\rangle_{C_1}|s_2\rangle_{R_2}|t_2\rangle_{C_2}.
\end{eqnarray}

The core objective of this algorithm is to compute the product of each corresponding element pair from two matrices, thereby enabling the implementation of a quantum algorithm for the Hadamard product of matrices. Based on this, the algorithm first needs to filter out in the initial state that the product of the elements of two matrices is in the same row, that is, the term in the quantum state $|\Phi_0\rangle$ where the register $R_1$ and $R_2$ have the same state.
To this end, a $n$-qubit auxiliary $B_1$ with a state of $|0\rangle_{B_1}$ is introduced, and an operator is defined
\begin{eqnarray*}
W_{j}^{(y)} &=& P^{(y)}_{j} \otimes \sigma^{(x)}_{{B_1}^{(j)}} + (I_{j}-P^{(y)}_{j}) \otimes I_{{B_1}^{(j)}},
\end{eqnarray*}
where $P^{(y)}_j=|y_{j}\rangle_{R_1}|y_{j}\rangle_{R_2}  \;{ _{R_1}}\langle y_{j}|  {_{R_2}}\langle y_{j}| \;\; (y=0,1)$ is a projection operator acting on the $j$-th qubit of registers $R_1$ and $R_2$, $\sigma^{(x)}_{{B_1}^{(j)}}$ is a Pauli operator acting on the $j$-th qubit of the auxiliary register $B_1$, and $I_{{B_1}^{(j)}}$ is an identity operator acting on the $j$-th qubit of the auxiliary register $B_1$. By applying the operator $W_{R_1R_2B_1}^{(1)}=\prod_{j=1}^{n}W_{j}^{0}W_{j}^{1}$ (where $\prod$ denotes the tensor product) to $|\Phi_0\rangle |0\rangle_{B_1}$, the following result is obtained:
\begin{eqnarray}
	|\Phi_1\rangle &=& W^{(1)}_{R_1R_2B_1}  |\Phi_0\rangle|0\rangle_{B_1}\\\nonumber
	&=&\sum_{s=0}^{N-1}\sum_{t_1,t_2=0}^{M-1} a^{(1)}_{st_1}a^{(2)}_{st_2}|s\rangle_{R_1}|t_1\rangle_{C_1}|s\rangle_{R_2}|t_2\rangle_{C_2} \otimes |N-1\rangle_{B_1}+|g_1\rangle_{R_1C_1R_2C_2B_1}.
\end{eqnarray}

In the process of implementing the quantum state $|\Phi_1\rangle$, the algorithm needs to determine whether the states of the $j$-th qubits of registers $R_1$ and $R_2$ are the same. If the states of the $j$-th qubits of the two registers are identical, the $j$-th qubit of $B_1$ is flipped from $|0\rangle$ to $|1\rangle$. If the states of all $n$ qubits of $R_1$ and $R_2$ are the same, the overall state of the $n$ qubits in $B_1$ is transformed from $|0\rangle_{B_1}$ to $|N-1\rangle_{B_1}$. For the product terms of elements that do not belong to the same row in the two matrices, these unnecessary information terms (which are not required for subsequent operations) are uniformly represented by $|g_1\rangle_{R_1C_1R_2C_2B_1}$.

In the process described above, we identified terms representing products of elements from the same row of the two matrices. Building on this, we will now isolate terms where the products of elements from the two matrices belong to the same column, specifically, the terms in quantum state $|\Phi_1\rangle$ where registers $C_1$ and $C_2$ exhibit identical states.
To achieve this, we introduce an $m$-qubit auxiliary register $B_2$ initialized in the $|0\rangle_{B_2}$ state and define the operator:
\begin{eqnarray}
W_{j}^{(y)} &=& P^{(y)}_{j} \otimes \sigma^{(x)}_{{B_2}^{(j)}} + (I_{j}-P^{(y)}_{j}) \otimes I_{{B_2}^{(j)}},
\end{eqnarray}
where $P^{(y)}_j=|y_{j}\rangle_{C_1}|y_{j}\rangle_{C_2}  \;{ _{C_1}}\langle y_{j}|  {_{C_2}}\langle y_{j}| \;\; (y=0,1)$ serves as the projection operator. When applying the operator $W_{C_1C_2B_2}^{(2)}=\prod_{j=1}^{m}W_{j}^{0}W_{j}^{1}$ to $|\Phi_1\rangle |0\rangle_{B_2}$, we obtain:
\begin{eqnarray}
|\Phi_2\rangle &=& W^{(2)}_{C_1C_2B_2}|\Phi_1\rangle |0\rangle_{B_2} \\\nonumber
&=&\sum_{s=0}^{N-1}\sum_{t=0}^{M-1} a^{(1)}_{st}a^{(2)}_{st}|s\rangle_{R_1}|t\rangle_{C_1}|s\rangle_{R_2}|t\rangle_{C_2}  \otimes|N-1\rangle_{B_1} |M-1\rangle_{B_2} +|g_2\rangle_{R_1C_1R_2C_2B_1B_2} .
\end{eqnarray}
Terms representing element products that belong to neither the same row nor the same column in the two matrices constitute irrelevant information for this algorithm. These superfluous terms are collectively denoted by $|g_2\rangle_{R_1C_1R_2C_2B_1B_2}$.

The information corresponding to the Hadamard product of the two matrices is now contained within the first term of $|\Phi_2\rangle$. To effectively prevent the intermixing of useful and extraneous information terms in subsequent computational processes, we introduce a 1-qubit auxiliary register $B_3$ initialized in the $|0\rangle_{B_3}$ state, along with a projection operator $P_{B_1B_2} = |N-1\rangle_{B_1}|M-1\rangle_{B_2} \; {}_{B_1}\langle N-1| \; {}_{B_2}\langle M-1|$. This setup enables $B_3$ to perform distinctive tagging operations on the useful and useless information terms within $|\Phi_2\rangle$.
To achieve this, we construct the controlled operator $W^{(3)}_{B_1B_2B_3} = P_{B_1B_2} \otimes \sigma^{(x)}_{B_3} + (I_{B_1B_2} - P_{B_1B_2}) \otimes I_{B_3}$, which acts on $|\Phi_2\rangle|0\rangle_{B_3}$ to yield:
\begin{eqnarray}
|\Phi_3\rangle &=& W^{(3)}_{B_1B_2B_3}|\Phi_2\rangle |0\rangle_{B_3} \\\nonumber
&=& \sum_{s=0}^{N-1}\sum_{t=0}^{M-1} a^{(1)}_{st}a^{(2)}_{st}|s\rangle_{R_1}|t\rangle_{C_1}|s\rangle_{R_2}|t\rangle_{C_2} |N-1\rangle_{B_1} |M-1\rangle_{B_2}|1\rangle_{B_3} + |g_2\rangle_{R_1C_1R_2C_2B_1B_2}|0\rangle_{B_3}.
\end{eqnarray}

The controlled operator $W^{(3)}_{B_1B_2B_3}$, featuring $n+m$ control qubits, can be implemented using $O(n+m)$ Toffoli gates \cite{KShV}. Consequently, the circuit depth required to generate $|\Phi_2\rangle$ is $O(n+m) = O(\max\{n,m\})$, and the overall circuit complexity is likewise $O(n+m) = O(\max\{n,m\})$.

In the term of the quantum state $|\Phi_4\rangle$ marked by $|1\rangle_{B_3}$, due to the different states of $R_2$ and $C_2$, this state difference directly leads to the inability to perform the merging operation on the useful information terms. Based on this, Hadamard gates are applied to $R_2$ and $C_2$ to transform their states into superposition states.

By applying the Hadamard operator $W^{(4)}_{R_2C_2B_1B_2}=H^{\otimes (n+m)}$ to $|\Phi_3\rangle$, obtain:
\begin{eqnarray}
|\Phi_4\rangle &=&W^{(4)}_{R_2C_2B_1B_2}|\Phi_3\rangle\\\nonumber
& =&\frac{1}{2^{{(n+m)}/2}}
\sum_{s=0}^{N-1}\sum_{t=0}^{M-1} a^{(1)}_{st}a^{(2)}_{st}|s\rangle_{R_1}|t\rangle_{C_1}|0\rangle_{R_2}|0\rangle_{C_2} |N-1\rangle_{B_1} |M-1\rangle_{B_2}|1\rangle_{B_3}+|g_3\rangle_{R_1C_1R_2C_2B_1B_2B_3}
\end{eqnarray}
Since a new extraneous information term $|g_3\rangle_{R_1C_1R_2C_2B_1B_2B_3}$ was generated in the previous step, we introduce a 1-qubit auxiliary register $B_4$ in the $|0\rangle_{B_4}$ state along with the projection operator:
$
P_{R_2C_2B_1B_2B_3}= |0\rangle_{R_2}|0\rangle_{C_2}|N-1\rangle_{B_1}|M-1\rangle_{B_2}|1\rangle_{B_3}\;{_{R_2}}\langle 0| {_{C_2}}\langle 0| _{B_1}\langle N-1|_{B_2}\langle M-1| _{B_3}\langle 1|
$
to re-label the useful and useless information terms appropriately.

We then construct the controlled operator:
$
W^{(5)}_{R_2C_2B_1B_2B_3B_4}=P_{R_2C_2B_1B_2B_3} \otimes \sigma^{(x)}_{B_4}+
(I_{R_2C_2B_1B_2B_3} -P_{R_2C_2B_1B_2B_3} )\otimes I_{B_4}
$
, which acts on $|\Phi_4\rangle \otimes |0\rangle_{B_4}$ to yield:
\begin{eqnarray}
	|\Phi_5\rangle &=&W^{(5)}_{R_2C_2B_1B_2B_3B_4}|\Phi_4\rangle |0\rangle_{B_4}
\\\nonumber
&=&\frac{1}{2^{{(n+m)}/2}}
\sum_{s=0}^{N-1}\sum_{t=0}^{M-1} a^{(1)}_{st}a^{(2)}_{st}|s\rangle_{R_1}|t\rangle_{C_1}|0\rangle_{R_2}|0\rangle_{C_2}  \\\nonumber
&&|N-1\rangle_{B_1} |M-1\rangle_{B_2}|1\rangle_{B_3}|1\rangle_{B_4}+|g_3\rangle_{R_1C_1R_2C_2B_1B_2B_3}|0\rangle_{B_3}.
\end{eqnarray}
The control operator $W^{(5)}_{R_2C_2B_1B_2B_3B_4}$ with $2n+2m+1$ control qubits can be represented by $O(2n+2m+1)$ Toffoli gate \cite{KShV}. Therefore, the depth of the $|\Phi_5\rangle$ circuit is calculated as $O(2n+2m+1)=O(max\{n,m\})$, and the complexity of the circuit is $O(max\{n,m\})$.

The information of the Hadamard product of two matrices is stored in the first term of $|\Phi_5\rangle$. By measuring the auxiliary $B_4$ with $|1\rangle_{B_4}$ and deleting the useless information terms while retaining the first term in the quantum state $|\Phi_5\rangle$ , obtain:
\begin{eqnarray}
|\Phi_6\rangle &=&|\Phi_{out}\rangle|0\rangle_{R_2}|0\rangle_{C_2}
|N-1\rangle_{B_1} |M-1\rangle_{B_2}|1\rangle_{B_3},
\end{eqnarray}
where $|\Phi_{out}\rangle =G^{-1}\sum_{s=0}^{N-1}\sum_{t=0}^{M-1} a^{(1)}_{st}a^{(2)}_{st}|s\rangle_{R_1}|t\rangle_{C_1}$, and
the normalization term $G=(\sum_{s}  \sum_{t} |{a^{(1)}_{st}a^{(2)}_{st}}|^2)^{1/2}$, then the success probability of the above measurement is {$\frac {G^2}{2^{(n+m)}}$}.

As discussed, the overall computational depth of the algorithm is determined by the operators $W^{(3)}_{B_1B_2B_3}$ and $W^{(5)}_{R_2C_2B_1B_2B_3B_4}$. Thus, the total computational complexity of the algorithm is $O(\max\{n,m\})$.

\subsection{ Kronecker Product of Matrices}

In this section, let $A^{(1)}$ be an $N \times M$ matrix and $A^{(2)}$ be an $M \times K$ matrix, where the elements of matrix $A^{(i)}$ are denoted as $A^{(i)} = (a^{(i)}_{s_i t_i})$ with $i = 1, 2$ ($N = 2^n$, $M = 2^m$, $K = 2^k$; $n$, $m$, and $k$ are positive integers). The circuit diagram of the quantum algorithm for computing the Kronecker product of the two matrices is presented in Figure 2.

\begin{figure}[h]
\centerline{\includegraphics[width=0.5\textwidth]{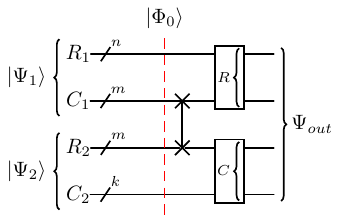}}
\caption{The quantum circuit diagram for the Kronecker product}
\end{figure}

To implement this quantum algorithm, we first introduce one $n$-qubit register $R_1$, two $m$-qubit registers $C_1$ and $R_2$, and one $k$-qubit register $C_2$. Register $R_1$ and $C_1$ enumerate the rows and columns of matrix $A^{(1)}$, while $R_2$ and $C_2$ enumerate the rows and columns of matrix $A^{(2)}$. The elements of the two matrices are encoded into pure states respectively as:
\begin{eqnarray}
&&|\Psi_1\rangle = \sum_{s_1=0}^{N-1}\sum_{t_1=0}^{M-1}a^{(1)}_{s_1t_1}|s_1\rangle_{R_1}|t_1\rangle_{C_1},\\\nonumber
&&|\Psi_2\rangle = \sum_{s_2=0}^{M-1}\sum_{t_2=0}^{K-1}a^{(2)}_{s_2t_2}|s_2\rangle_{R_2}|t_2\rangle_{C_2},\\\nonumber
\end{eqnarray}
where \(\sum_{s_i t_i}|a^{(i)}_{s_i t_i}|^2=1,\;\;i=1,2.\)

The initial state of the entire system is constructed by taking the tensor product of the above two pure states:
\begin{eqnarray}
|\Phi_0\rangle &= & |\Psi_1\rangle \otimes |\Psi_2\rangle  \\\nonumber
&=&\sum_{s_1=0}^{N-1} \sum_{t_1,s_2=0}^{M-1}\sum_{t_2=0}^{K-1} 
 a^{(1)}_{s_1 t_1} a^{(2)}_{s_2 t_2} |s_1\rangle_{R_1} |t_1\rangle_{C_1}  |s_2\rangle_{R_2} |t_2\rangle_{C_2}.
\end{eqnarray}

To store the row information of the two matrices in registers \( R_1 \) and \( C_1 \), and their column information in registers \( R_2 \) and \( C_2 \), we achieve this by using a controlled \( SWAP \) operation to exchange the information states of registers \( C_1 \) and \( R_2 \). For this purpose, we apply the controlled operator \( SWAP_{C_1R_2} \) to \( |\Phi_0\rangle \), resulting in:
\begin{eqnarray}
|\Phi_1\rangle &=& W^{(1)}_{C_1R_2} |\Phi_0\rangle\\\nonumber
&=&\sum_{s_1=0}^{N-1} \sum_{s_2,t_1=0}^{M-1}\sum_{t_2=0}^{K-1} 
 a^{(1)}_{s_1 t_1} a^{(2)}_{s_2 t_2} |s_1\rangle_{R_1} |s_2\rangle_{C_1}  |t_1\rangle_{R_2} |t_2\rangle_{C_2} \\\nonumber
&=& \sum_{s_1=0}^{N-1} \sum_{s_2,t_1=0}^{M-1}\sum_{t_2=0}^{K-1}
 a^{(1)}_{s_1,t_1} a^{(2)}_{s_2,t_2}|Ms_1+s_2\rangle_{R} |Kt_1+t_2\rangle_{C}
\end{eqnarray}

Here, since \( |s_1\rangle_{R_1} |s_2\rangle_{C_1}=|Ms_1+s_2\rangle_{R} \) and \( |t_1\rangle_{R_2} |t_2\rangle_{C_2}=|Kt_1+t_2\rangle_{C} \), we combine registers to more clearly represent the target matrix:  
 Registers \( R_1 \) and \( C_1 \) are merged into a new \( (n+m) \)-qubit register \( R \), which is used to enumerate the rows of the new matrix.  
 Registers \( R_2 \) and \( C_2 \) are merged into a new \( (m+k) \)-qubit register \( C \), which is used to enumerate the columns of the new matrix.  
Since the \( SWAP \) gate acts on distinct qubit pairs of registers \( C_1 \) and \( R_2 \), the entire algorithm requires \( m \) \( SWAP \) gates, and the computational depth is \( O(1) \).

\section{Elementary Column Transformations of Matrices}

This section presents the quantum algorithm for elementary column transformations of matrices, including two specific operations: adding one column of a matrix to another (column addition) and swapping two columns of a matrix (column swapping£©.

\subsection{Column addition}

In this section, let \( A = \{a_{ij}\} \) (where \( N=2^n \), \( M=2^m \)) be an \( N \times M \) matrix. The quantum circuit diagram for the algorithm that adds one column of matrix \( A \) to another column is shown in Figure 3.

\begin{figure}[h]
\centerline{\includegraphics[width=0.5\textwidth]{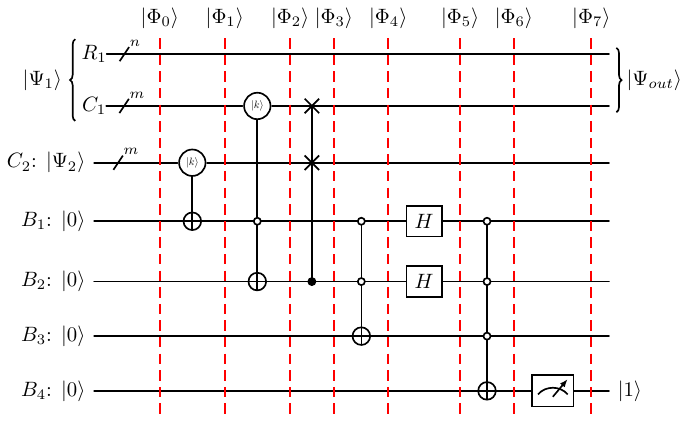}}
\caption{The quantum circuit diagram of column addition}
\end{figure}

To implement this quantum algorithm, we first introduce an \( n \)-qubit register \( R_1 \) and an \( m \)-qubit register \( C_1 \), which are used to index the rows and columns of matrix \( A \), respectively. The elements of matrix \( A \) are then encoded into a pure quantum state, which is given by:  
\begin{eqnarray*}
|\Psi_1\rangle = \sum_{i=0}^{N-1}  \sum_{j=0}^{M-1}   a_{ij} |i\rangle_{R_1} |j\rangle_{C_1} ,\;\;~\sum_{i,j}|a_{ij}|^2=1.
\end{eqnarray*}

To more precisely specify the operation of adding one matrix column to another, "adding one column to another" here refers to adding the \((k+1)\)-th column of matrix \(A\) to the \((l+1)\)-th column, where \(0 \leq l, k \leq M-1\). 
To implement the operation of adding the \((k+1)\)-th column to the \((l+1)\)-th column of the matrix, we introduce an auxiliary column vector of dimension \(2^m\). This auxiliary vector has elements \(\frac{1}{\sqrt{2}}\) only in the \((l+1)\)-th and \((k+1)\)-th rows, with all other elements being zero. We then introduce an \(m\)-qubit register \(C_2\) to encode this auxiliary vector as a pure state:
\begin{eqnarray*}
|\Psi_2\rangle = \frac{1}{\sqrt{2}}(|k\rangle_{C_2} +|l\rangle_{C_2}).
\end{eqnarray*}

The initial state of the entire system is constructed by taking the tensor product of the above two pure states:
\begin{eqnarray}
|\Phi_0\rangle &=&  |\Psi_1\rangle \otimes |\Psi_2\rangle\\\nonumber
&=&\frac{1}{\sqrt{2}}\Big(\sum_{i=0}^{N-1} \sum_{j=0}^{M-1}  a_{ij}  |i\rangle_{R_1} |j\rangle_{C_1}  |k\rangle_{C_2}  
+\sum_{i=0}^{N-1} \sum_{j=0}^{M-1} a_{ij}  |i\rangle_{R_1} |j\rangle_{C_1}  |l\rangle_{C_2} \Big).
\end{eqnarray}

To implement the operation of adding the \((k+1)\)-th column to the \((l+1)\)-th column in matrix \(A\), we first need to preserve all terms where register \(C_2\) remains in the state \(|k\rangle_{C_2}\) while simultaneously retaining the elements from the \((k+1)\)-th column in those terms where \(C_2\) is in the state \(|l\rangle_{C_2}\). This is achieved through operations (17) and (18) as follows.
First, we introduce a 1-qubit auxiliary register \(B_1\) in the state \(|0\rangle_{B_1}\) and a projection operator \(P_{C_2} = |k\rangle_{C_2} \, {}_{C_2}\langle k|\). We then apply the controlled operator:
\begin{eqnarray*}
W^{(1)}_{C_2B_1} = P_{C_2} \otimes \sigma^{(x)}_{B_1} + (I_{C_2} - P_{C_2}) \otimes I_{B_1}
\end{eqnarray*}
acting on \(C_2\) and \(B_1\) respectively, resulting in:
\begin{eqnarray}
|\Phi_1\rangle &=& W^{(1)}_{C_2B_1} |\Phi_0\rangle \otimes |0\rangle_{B_1}\\\nonumber
&=& \frac{1}{\sqrt{2}} \Big( \sum_{i=0}^{N-1} \sum_{j=0}^{M-1} a_{ij} |i\rangle_{R_1} |j\rangle_{C_1} |k\rangle_{C_2} |1\rangle_{B_1} 
 + \sum_{i=0}^{N-1} \sum_{j=0}^{M-1} a_{ij} |i\rangle_{R_1} |j\rangle_{C_1} |l\rangle_{C_2} |0\rangle_{B_1} \Big).
\end{eqnarray}

The controlled operator \(W^{(1)}_{C_2B_1}\) with \(m\) control qubits can be implemented using \(O(m)\) Toffoli gates \cite{KShV}. Consequently, the circuit depth required to generate \(|\Phi_1\rangle\) is \(O(m)\), and the overall circuit complexity is also \(O(m)\).

Next, a 1-qubit auxiliary register \( B_2 \) initialized to the state \( |0\rangle_{B_2} \) and a projection operator \( P_{C_1B_1} = |k\rangle_{C_1} |0\rangle_{B_1} \; {_{C_1}}\langle k| \; {_{B_1}}\langle 0| \) are introduced. Subsequently, the control operator  
\begin{eqnarray*}
W^{(2)}_{C_1B_1B_2} &=& P_{C_1B_1} \otimes \sigma^{(x)}_{B_2} + (I_{C_1B_1} - P_{C_1B_1}) \otimes I_{B_2}
\end{eqnarray*}
is applied to the state \( |\Phi_1\rangle |0\rangle_{B_2} \). This operation extracts the elements of the \( (k+1) \)-th column from the terms marked by \( |0\rangle_{B_1} \), resulting in the state:  
\begin{eqnarray}
|\Phi_2\rangle &=& W^{(2)}_{C_1B_1B_2} \left( |\Phi_1\rangle \otimes |0\rangle_{B_2} \right) \\\nonumber
&=& \frac{1}{\sqrt{2}} \Bigg( \sum_{i=0}^{N-1} \sum_{j=0}^{M-1} a_{ij} |i\rangle_{R_1} |j\rangle_{C_1} |k\rangle_{C_2} |1\rangle_{B_1} |0\rangle_{B_2} + \sum_{i=0}^{N-1} a_{ik} |i\rangle_{R_1} |k\rangle_{C_1} |l\rangle_{C_2} |0\rangle_{B_1} |1\rangle_{B_2} \Bigg) \\\nonumber
&& + |g_1\rangle_{R_1C_1C_2} |0\rangle_{B_1} |0\rangle_{B_2}.
\end{eqnarray}
Here, the terms marked with \( |0\rangle_{B_1} |0\rangle_{B_2} \) represent irrelevant information that is not required by the algorithm, denoted by \( |g_1\rangle \), while the remaining terms correspond to the useful information needed for the algorithm.  
The depth of the $|\Phi_2\rangle$ circuit is calculated as $O(m)$, and the complexity of the circuit is $O(m)$.

Now, to move the elements of the \( (k+1) \)-th column (marked by \( |1\rangle_{B_2} \)) to the \( (l+1) \)-th column, a controlled SWAP gate is employed here. This gate swaps the states of registers \( C_1 \) and \( C_2 \), with \( |1\rangle_{B_2} \) serving as the control qubit. We first construct the control operator:  
\begin{eqnarray*}
W^{(3)}_{C_1C_2B_2} = SWAP_{C_1,C_2} \otimes |1\rangle_{B_2} \; {_{B_2}}\langle 1| + I_{C_1C_2} \otimes |0\rangle_{B_2} \; {_{B_2}}\langle 0|
\end{eqnarray*}  
We then apply this operator to \( |\Phi_2\rangle \), swapping the states of \( C_1 \) and \( C_2 \) to obtain the state:  
\begin{eqnarray}
|\Phi_3\rangle &=& W^{(3)}_{C_1C_2B_2} |\Phi_2\rangle \\\nonumber
&=& \frac{1}{\sqrt{2}} \Bigg( \sum_{i=0}^{N-1} \sum_{j=0}^{M-1} a_{ij} |i\rangle_{R_1} |j\rangle_{C_1} |k\rangle_{C_2} |1\rangle_{B_1} |0\rangle_{B_2} + \sum_{i=0}^{N-1} a_{ik} |i\rangle_{R_1} |l\rangle_{C_1} |k\rangle_{C_2} |0\rangle_{B_1} |1\rangle_{B_2} \Bigg) \\\nonumber
&& + |g_1\rangle_{R_1C_1C_2} |0\rangle_{B_1} |0\rangle_{B_2}.
\end{eqnarray}  

The swap operations within the control operator \( W^{(3)}_{C_1C_2B_2} \) share a single control qubit. Comprising \( m \) C-SWAP gates (one for each pair of qubits across registers \( C_1 \) and \( C_2 \)), these gates target distinct qubit pairs. As a result, they can be implemented concurrently. For this reason, the depth of this operator is \( O(1) \), and its complexity is \( O(m) \).

To prevent useful information terms and irrelevant information terms from mixing in subsequent calculations (which would compromise the final measurement results), an auxiliary qubit is introduced to label these two types of terms separately. Here, we introduce a 1-qubit auxiliary register \( B_3 \) initialized to the state \( |0\rangle_{B_3} \), along with a projection operator \( P_{B_1B_2} = |00\rangle_{B_1B_2} \; {_{B_1B_2}}\langle 00| \). We then apply the control operator  
\begin{eqnarray*}
W^{(4)}_{B_1B_2B_3} = P_{B_1B_2} \otimes \sigma^{(x)}_{B_3} + (I_{B_1B_2} - P_{B_1B_2}) \otimes I_{B_3}
\end{eqnarray*}  
to the state \( |\Phi_3\rangle |0\rangle_{B_3} \), yielding the state:  
\begin{eqnarray}
|\Phi_4\rangle &=& W^{(4)}_{B_1B_2B_3} \left( |\Phi_3\rangle \otimes |0\rangle_{B_3} \right) \\\nonumber
&=& \frac{1}{\sqrt{2}} \Bigg( \sum_{i=0}^{N-1} \sum_{j=0}^{M-1} a_{ij} |i\rangle_{R_1} |j\rangle_{C_1} |k\rangle_{C_2} |1\rangle_{B_1} |0\rangle_{B_2} \\\nonumber
&& \quad + \sum_{i=0}^{N-1} a_{ik} |i\rangle_{R_1} |l\rangle_{C_1} |k\rangle_{C_2} |0\rangle_{B_1} |1\rangle_{B_2} \Bigg) \otimes |0\rangle_{B_3} + |g_1\rangle_{R_1C_1C_2B_1B_2} \otimes |1\rangle_{B_3}.
\end{eqnarray}

Next, we apply the Hadamard transform \( W^{(5)}_{B_1B_2} = H^{\otimes 2} \) to registers \( B_1 \) and \( B_2 \), resulting in the state:  
\begin{eqnarray}
|\Phi_5\rangle &=& W^{(5)}_{B_1B_2} |\Phi_4\rangle  \\\nonumber
&=& \frac{1}{(\sqrt{2})^{3}} \Bigg( \sum_{i=0}^{N-1} \sum_{j=0}^{M-1} a_{ij} |i\rangle_{R_1} |j\rangle_{C_1} + \sum_{i=0}^{N-1} a_{ik} |i\rangle_{R_1} |l\rangle_{C_1} \Bigg) |k\rangle_{C_2} |0\rangle_{B_1} |0\rangle_{B_2} |0\rangle_{B_3} \\\nonumber
&& + |g_2\rangle_{R_1C_1C_2B_1B_2B_3}.
\end{eqnarray}  

To label the newly generated irrelevant information terms, we introduce a 1-qubit auxiliary register \( B_4 \) initialized to the state \( |0\rangle_{B_4} \), along with a projection operator \( P_{B_1B_2B_3} = |000\rangle_{B_1B_2B_3} \; {_{B_1B_2B_3}}\langle 000| \). We then apply the control operator  
\begin{eqnarray*}
W^{(6)}_{B_1B_2B_3B_4} = P_{B_1B_2B_3} \otimes \sigma^{(x)}_{B_4} + (I_{B_1B_2B_3} - P_{B_1B_2B_3}) \otimes I_{B_4}
\end{eqnarray*}  
to the state \( |\Phi_5\rangle |0\rangle_{B_4} \), yielding the state:  
\begin{eqnarray}
|\Phi_6\rangle &=& W^{(6)}_{B_1B_2B_3B_4} \left( |\Phi_5\rangle \otimes |0\rangle_{B_4} \right) \\\nonumber
&=& \frac{1}{(\sqrt{2})^{3}} \Bigg( \sum_{i=0}^{N-1} \sum_{j=0}^{M-1} a_{ij} |i\rangle_{R_1} |j\rangle_{C_1}+ \sum_{i=0}^{N-1} a_{ik} |i\rangle_{R_1} |l\rangle_{C_1} \Bigg) |k\rangle_{C_2} |0\rangle_{B_1} |0\rangle_{B_2} |0\rangle_{B_3} |1\rangle_{B_4} \\\nonumber
&& + |g_2\rangle_{R_1C_1C_2B_1B_2B_3} \otimes |0\rangle_{B_4}.
\end{eqnarray}

To label the newly generated irrelevant information terms, we introduce a 1-qubit auxiliary register \( B_4 \) initialized to the state \( |0\rangle_{B_4} \) and a projection operator \( P_{B_1B_2B_3} = |000\rangle_{B_1B_2B_3} \langle 000| \). We then apply the control operator  
\begin{eqnarray*}
W^{(6)}_{B_1B_2B_3B_4} = P_{B_1B_2B_3} \otimes \sigma^{(x)}_{B_4} + (I_{B_1B_2B_3} - P_{B_1B_2B_3}) \otimes I_{B_4}
\end{eqnarray*}  
to the state \( |\Phi_5\rangle \otimes |0\rangle_{B_4} \), resulting in the state:  
\begin{eqnarray}
|\Phi_6\rangle &=& W^{(6)}_{B_1B_2B_3B_4} \left( |\Phi_5\rangle \otimes |0\rangle_{B_4} \right) \\\nonumber
&=& \frac{1}{(\sqrt{2})^3} \Bigg( \sum_{i=0}^{N-1} \sum_{j=0}^{M-1} a_{ij} |i\rangle_{R_1} |j\rangle_{C_1} + \sum_{i=0}^{N-1} a_{ik} |i\rangle_{R_1} |l\rangle_{C_1} \Bigg) |k\rangle_{C_2} |0\rangle_{B_1} |0\rangle_{B_2} |0\rangle_{B_3} |1\rangle_{B_4} \\\nonumber
&& + |g_2\rangle_{R_1C_1C_2B_1B_2B_3} \otimes |0\rangle_{B_4}.
\end{eqnarray}

By measuring the auxiliary register \( B_4 \) for the state \( |1\rangle_{B_4} \), we can eliminate irrelevant information terms while retaining only useful information terms. This measurement yields the state:  
\begin{eqnarray}
|\Phi_7\rangle &=& |\Psi_{out}\rangle \otimes |k\rangle_{C_2} \otimes |000\rangle_{B_1B_2B_3},
\end{eqnarray}  
where  
$$
|\Psi_{out}\rangle = \frac{1}{G} \Bigg( \sum_{i=0}^{N-1} \sum_{j \neq l} a_{ij} |i\rangle_{R_1} |j\rangle_{C_1} + \sum_{i=0}^{N-1} (a_{il} + a_{ik}) |i\rangle_{R_1} |l\rangle_{C_1} \Bigg)
$$  
and the normalization factor \( G \) is given by  
$$
G = \Bigg( \sum_{i=0}^{N-1} \sum_{j \neq l} |a_{ij}|^2 + \sum_{i=0}^{N-1} |a_{il} + a_{ik}|^2 \Bigg)^{1/2}.
$$  

The success probability of this measurement in the algorithm is \( \frac{G^2}{8} \), and this probability depends solely on the elements \( a_{ij} \) of the matrix. As discussed above, the overall computational complexity of the algorithm is determined by the operators \( W^{(1)}_{C_2B_1} \), \( W^{(2)}_{C_1B_1B_2} \), and \( W^{(3)}_{C_1C_2B_2} \); thus, the overall computational complexity of the algorithm is \( O(m) \).

\textbf{Remark 1.} Compared with the quantum algorithm that first performs a matrix transpose and then adds one row to another (as an indirect way to achieve the addition of one column to another of the matrix), the quantum algorithm that directly implements the addition of one column to another of the matrix not only retains the same computational complexity as the row addition algorithm in \cite{LTLF}, but more importantly, completely eliminates the additional quantum resource consumption incurred by the transpose operation.

\subsection{Column swapping}

In this section, let \( A = \{a_{ij}\} \) denote an \( N \times M \) matrix, where \( N = 2^n \) and \( M = 2^m \) (with \( n \) and \( m \) being positive integers). The circuit diagram of the quantum algorithm for computing the interchange of two columns of matrix \( A \) is presented in Figure 4.  
\begin{figure}[h]
\centerline{\includegraphics[width=0.5\textwidth]{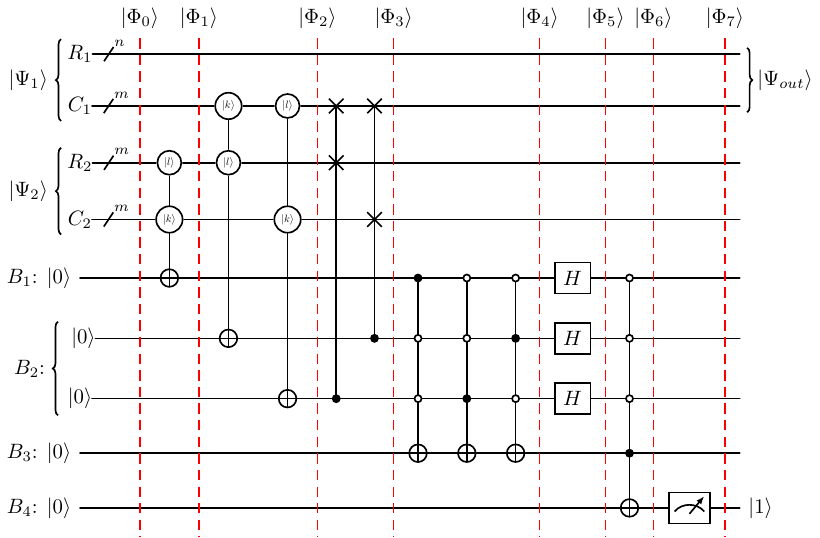}}
\caption{The quantum circuit diagram of column swapping}
\end{figure}

To implement this quantum algorithm, the $n$-qubit register $R_1$ and the $m$-qubit register $C_1$ are first introduced to enumerate the rows and columns of matrix $A$. Encode the elements of the matrix as pure states
\begin{eqnarray*}
|\Psi_1\rangle = \sum_{i=0}^{N-1}  \sum_{j=0}^{M-1}   a_{ij} |i\rangle_{R_1} |j\rangle_{C_1} ,\;\;~\sum_{ij}|a_{ij}|^2=1.
\end{eqnarray*}

To elaborate on the two-column interchange of a matrix in greater specificity, this operation refers to swapping the \((k+1)\)-th column and the \((l+1)\)-th column of matrix \(A\), where the indices satisfy \(0 \leq l, k \leq M-1\). To realize the interchange between the \((k+1)\)-th column and the \((l+1)\)-th column of the matrix, an auxiliary matrix of dimension \(M \times M\) is introduced herein. Within this auxiliary matrix, three specific elements take the value of \(\frac{1}{\sqrt{3}}\): these are the element at the \((l+1)\)-th row and \((k+1)\)-th column, the element at the \((k+1)\)-th row and \((k+1)\)-th column, and the element at the \((l+1)\)-th row and \((l+1)\)-th column. All other elements of the auxiliary matrix are zero.
Two \(m\)-qubit registers, denoted as \(R_2\) and \(C_2\), are introduced to encode the auxiliary matrix into a pure quantum state, which is defined as follows:
\begin{eqnarray*}
|\Psi_2\rangle = \frac{1}{\sqrt{3}}(|l\rangle_{R_2}|k\rangle_{C_2} + |k\rangle_{R_2}|k\rangle_{C_2} + |l\rangle_{R_2}|l\rangle_{C_2}).
\end{eqnarray*}

The initial state of the entire system is constructed by taking the tensor product of the above two pure states (i.e., \(|\Psi_1\rangle\) and \(|\Psi_2\rangle\)), which is expressed as:
\begin{eqnarray}
|\Phi_0\rangle &=&  |\Psi_1\rangle \otimes |\Psi_2\rangle\\\nonumber
&=&\frac{1}{\sqrt{3}}\Big(\sum_{i=0}^{N-1} \sum_{j=0}^{M-1}  a_{ij}  |i\rangle_{R_1} |j\rangle_{C_1}  |l\rangle_{R_2}|k\rangle_{C_2} \\\nonumber &&+\sum_{i=0}^{N-1} \sum_{j=0}^{M-1}  a_{ij}  |i\rangle_{R_1} |j\rangle_{C_1}|k\rangle_{R_2}|k\rangle_{C_2}\\\nonumber
&&+ \sum_{i=0}^{N-1} \sum_{j=0}^{M-1}  a_{ij}  |i\rangle_{R_1} |j\rangle_{C_1}  |l\rangle_{R_2}|l\rangle_{C_2} \Big).
\end{eqnarray}

To achieve the interchange between the \((k+1)\)-th column and the \((l+1)\)-th column of matrix \(A\), it is first necessary to retain all terms in register \(C_2\) except those corresponding to the \((k+1)\)-th column and the \((l+1)\)-th column, with these retained terms maintaining the state \(|k\rangle_{C_2}\). Additionally, the elements of the \((l+1)\)-th column should be retained in the terms where register \(C_2\) is in the state \(|k\rangle_{C_2}\), and the elements of the \((k+1)\)-th column should be retained in the terms where \(C_2\) is in the state \(|l\rangle_{C_2}\). This objective is accomplished through the operations described in equations (25) and (26) below.
First, a 1-qubit auxiliary register \(B_1\) initialized in the state \(|0\rangle_{B_1}\) and a projection operator \(P_{R_2C_2} = |l\rangle_{R_2}|k\rangle_{C_2} \langle l|_{R_2} \langle k|_{C_2}\) are introduced. Subsequently, a control operator is applied, which is defined as:
\begin{eqnarray*}
W^{(1)}_{R_2C_2B_1} = P_{R_2C_2} \otimes \sigma^{(x)}_{B_1} + (I_{R_2C_2}-P_{R_2C_2}) \otimes I_{B_1}
\end{eqnarray*}
This control operator acts on registers \(C_2\) and \(B_1\), resulting in the state \(|\Phi_1\rangle\) expressed as follows:
\begin{eqnarray}
|\Phi_1\rangle &=& W^{(1)}{R_2C_2B_1} |\Phi_0\rangle \otimes |0\rangle{B_1} \\\nonumber
&=& \frac{1}{\sqrt{3}} \sum_{i=0}^{N-1} \sum_{j=0}^{M-1} a_{ij} |i\rangle_{R_1} |j\rangle_{C_1} |l\rangle_{R_2}|k\rangle_{C_2} |1\rangle_{B_1} \\\nonumber
&& + \frac{1}{\sqrt{3}} \Bigg( \sum_{i=0}^{N-1} \sum_{j=0}^{M-1} a_{ij} |i\rangle_{R_1} |j\rangle_{C_1} |k\rangle_{R_2}|k\rangle_{C_2} \\\nonumber
&& + \sum_{i=0}^{N-1} \sum_{j=0}^{M-1} a_{ij} |i\rangle_{R_1} |j\rangle_{C_1} |l\rangle_{R_2}|l\rangle_{C_2} \Bigg) |0\rangle_{B_1}.
\end{eqnarray}
The circuit depth required to generate the state \(|\Phi_1\rangle\) is calculated as \(O(m) = O(\log M)\).

Next, a 2-qubit auxiliary register \(B_2\) initialized in the state \(|00\rangle_{B_2}\) is introduced, along with two projection operators: \(P_{C_1R_2} = |k\rangle_{C_1} |l\rangle_{R_2} \langle k|_{C_1} \langle l|_{R_2}\) and \(P_{C_1C_2} = |l\rangle_{C_1} |k\rangle_{C_2} \langle l|_{C_1} \langle k|_{C_2}\). Subsequently, the following two operators are applied:
\begin{eqnarray*}
W^{(1)} &=& P_{C_1R_2} \otimes \sigma^{(x)}{B_2^1} + (I{C_1R_2} - P_{C_1R_2}) \otimes I_{B_2^1}, \\\nonumber
W^{(2)} &=& P_{C_1C_2} \otimes \sigma^{(x)}{B_2^2} + (I{C_1C_2} - P_{C_1C_2}) \otimes I_{B_2^1},
\end{eqnarray*}
where \(I_{B_2^1}\) denotes the identity operator acting on the first qubit of \(B_2\), and \(\sigma^{(x)}_{B_2^1}\), \(\sigma^{(x)}_{B_2^2}\) represent the Pauli-X operators acting on the first and second qubits of \(B_2\), respectively.
A control operator \(W^{(2)}_{C_1R_2C_2B_2}\) is constructed as the product of \(W^{(1)}\) and \(W^{(2)}\), i.e., \(W^{(2)}_{C_1R_2C_2B_2} = W^{(1)}W^{(2)}\). This control operator acts on the combined state \(|\Phi_1\rangle |00\rangle_{B_2}\), leading to the state \(|\Phi_2\rangle\) which is expressed as:
\begin{eqnarray}
|\Phi_2\rangle &=&W^{(2)}_{C_1R_2C_2B_2} |\Phi_1\rangle \otimes |00\rangle_{B_2}\\\nonumber
 &=&\frac{1}{\sqrt{3}}\Big(\sum_{i=0}^{N-1} \sum_{j\neq l,k}  a_{ij}  |i\rangle_{R_1} |j\rangle_{C_1}  |l\rangle_{R_2}|k\rangle_{C_2} |1\rangle_{B_1} |00\rangle_{B_2} \\\nonumber
 &&
+\sum_{i=0}^{N-1} a_{il}  |i\rangle_{R_1} |l\rangle_{C_1}  |k\rangle_{R_2}|k\rangle_{C_2} |0\rangle_{B_1} |01\rangle_{B_2} \\\nonumber
&&
+\sum_{i=0}^{N-1} a_{ik}  |i\rangle_{R_1} |k\rangle_{C_1}  |l\rangle_{R_2}|l\rangle_{C_2}  |0\rangle_{B_1} |10\rangle_{B_2}\Big)\\\nonumber
&&+|g_1\rangle_{R_1C_1R_2C_2B_1B_2} .
\end{eqnarray}

In this algorithm, the term \(|g_1\rangle_{R_1C_1R_2C_2B_1B_2}\) (abbreviated as \(|g_1\rangle\)) is used to represent all unnecessary and irrelevant information items, while the remaining terms correspond to the useful information items required by the algorithm. The circuit depth needed to generate the state \(|\Phi_2\rangle\) is calculated as \(O(m) = O(\log M)\).

Now, to replace the element in the \((k+1)\)-th column (marked by the state \(|10\rangle_{B_2}\)) with the element in the \((l+1)\)-th column, and conversely replace the element in the \((l+1)\)-th column (marked by the state \(|01\rangle_{B_2}\)) with the element in the \((k+1)\)-th column, a controlled SWAP gate is employed. This gate is used to swap two sets of states: first, the states of registers \(C_1\) and \(C_2\), and second, the states of registers \(C_1\) and \(R_2\). The control bit for these swap operations is \(|1\rangle_{{B_2}^i}\) (where \(i = 1, 2\)), where \(|1\rangle_{{B_2}^1}\) indicates that the first qubit of auxiliary register \(B_2\) is in the state \(|1\rangle\), and \(|1\rangle_{B_2^2}\) indicates that the second qubit of \(B_2\) is in the state \(|1\rangle\).
Next, the following two operators are applied:
\begin{eqnarray*}
W^{(1)}&=&SWAP_{C_1,C_2} \otimes |1\rangle_{{B_2}^{1}} \;{_{{B_2}^{1}}}\langle 1|+
I_{C_1,C_2} \otimes |0\rangle_{{B_2}^{1}} \;{_{{B_2}^{1}}}\langle 0|,\\\nonumber
W^{(2)}&=&SWAP_{C_1,R_2} \otimes |1\rangle_{{B_2}^{2}} \;{_{{B_2}^{2}}}\langle 1|+
I_{C_1,R_2} \otimes |0\rangle_{{B_2}^{2}} \;{_{{B_2}^{2}}}\langle 0|
\end{eqnarray*}
 A control operator \(W^{(3)}_{C_1R_2C_2B_2}\) is then constructed as the product of \(W^{(1)}\) and \(W^{(2)}\), i.e., \(W^{(3)}_{C_1R_2C_2B_2} = W^{(2)}W^{(1)}\). This control operator acts on the state \(|\Phi_2\rangle\), resulting in the state \(|\Phi_3\rangle\) which is expressed as:
\begin{eqnarray}
|\Phi_3\rangle &=&W^{(3)}_{C_1R_2C_2B_2}|\Phi_2\rangle\\\nonumber
 &=&\frac{1}{\sqrt{3}}\Big(\sum_{i=0}^{N-1} \sum_{j\neq l,k}  a_{ij}  |i\rangle_{R_1} |j\rangle_{C_1}  |l\rangle_{R_2}|k\rangle_{C_2} |1\rangle_{B_1} |00\rangle_{B_2} \\\nonumber
 &&
+\sum_{i=0}^{N-1} a_{il}  |i\rangle_{R_1} |k\rangle_{C_1}  |l\rangle_{R_2}|k\rangle_{C_2} |0\rangle_{B_1} |01\rangle_{B_2} \\\nonumber
&&
+\sum_{i=0}^{N-1} a_{ik}  |i\rangle_{R_1} |l\rangle_{C_1}  |l\rangle_{R_2}|k\rangle_{C_2}  |0\rangle_{B_1} |10\rangle_{B_2}\Big)\\\nonumber
&&+|g_2\rangle_{R_1C_1R_2C_2B_1B_2}.
\end{eqnarray}
The circuit depth of the control operator \(W^{(3)}_{C_1R_2C_2B_2}\) is \(O(1)\). Notably, the swap operations within \(W^{(3)}_{C_1R_2C_2B_2}\) share a common control mechanism and consist of \(2m\) C-SWAP gates, which act on distinct pairs of qubits. Due to this structure, these C-SWAP gates can be applied simultaneously. Consequently, the operator exhibits a circuit depth of \(O(1)\) and a circuit complexity of \(O(m)\).

To avoid the mixing of useful information items and useless information items in subsequent calculation processes, which would otherwise interfere with the final measurement results, auxiliary marking is introduced. This marking mechanism is designed to separately label useful information items and useless information items. A 1-qubit auxiliary register \(B_3\) initialized in the state \(|0\rangle_{B_3}\) is introduced, and an operator \(V^{(i)}\) is defined as follows: \(V^{(i)} = P^{(i)}_{B_1B_2} \otimes \sigma^{(x)}_{B_3} + (I_{B_1B_2} - P_{B_1B_2}) \otimes I_{B_3}\). Here, \(P^{(i)}_{B_1B_2}\) represents a set of projection operators, specifically:
\begin{eqnarray*}
P^{(1)}_{B_1B_2}= |100\rangle_{B_1B_2}\langle 100|,\ \
P^{(2)}_{B_1B_2}= |001\rangle_{B_1B_2}\langle 001|,\ \
P^{(3)}_{B_1B_2}= |010\rangle_{B_1B_2} \langle 010|.
\end{eqnarray*}

Subsequently, a control operator \(W^{(4)}_{B_1B_2B_3}\) is constructed as the product of the \(V^{(i)}\) operators, i.e., \(W^{(4)}_{B_1B_2B_3} = V^{(1)}V^{(2)}V^{(3)}\). This control operator acts on the combined state \(|\Phi_3\rangle |0\rangle_{B_3}\), resulting in the state \(|\Phi_4\rangle\) which is expressed as:
\begin{eqnarray}
|\Phi_4\rangle&=& W^{(4)}_{B_1B_2B_3}|\Phi_3\rangle |0\rangle_{B_3} \\\nonumber
 &=&\frac{1}{\sqrt{3}}\Big(\sum_{i=0}^{N-1} \sum_{j\neq l,k}  a_{ij}  |i\rangle_{R_1} |j\rangle_{C_1}  |l\rangle_{R_2}|k\rangle_{C_2} |1\rangle_{B_1} |00\rangle_{B_2} \\\nonumber
 &&
+\sum_{i=0}^{N-1} a_{il}  |i\rangle_{R_1} |k\rangle_{C_1}  |l\rangle_{R_2}|k\rangle_{C_2} |0\rangle_{B_1} |01\rangle_{B_2} \\\nonumber
&&
+\sum_{i=0}^{N-1} a_{ik}  |i\rangle_{R_1} |l\rangle_{C_1}  |l\rangle_{R_2}|k\rangle_{C_2}  |0\rangle_{B_1} |10\rangle_{B_2}\Big)|1\rangle_{B_3} \\\nonumber
&&+|g_2\rangle_{R_1C_1R_2C_2B_1B_2}|0\rangle_{B_3}.
\end{eqnarray}
In this expression, useful information items are labeled with the state \(|1\rangle_{B_3}\) of the auxiliary register \(B_3\), while useless information items (represented by \(|g_2\rangle\)) are labeled with the state \(|0\rangle_{B_3}\), achieving effective separation of the two types of information.

Next, the Hadamard transform, denoted as \(W^{(5)}_{B_1B_2} = H^{\otimes 3}\), is applied to registers \(B_1\) and \(B_2\). 
After applying this Hadamard transform to the state \(|\Phi_4\rangle\), the resulting state \(|\Phi_5\rangle\) is expressed as follows:
\begin{eqnarray}
|\Phi_5\rangle &=&W^{(5)}_{B_1B_2} |\Phi_4\rangle  \\\nonumber
&=&\frac{1}{(\sqrt{2})^{3}}\frac{1}{\sqrt{3}}\Big(\sum_{i=0}^{N-1} \sum_{j\neq l,k}  a_{ij}  |i\rangle_{R_1} |j\rangle_{C_1}   
+\sum_{i=0}^{N-1} a_{il}  |i\rangle_{R_1} |k\rangle_{C_1}   
+\sum_{i=0}^{N-1} a_{ik}  |i\rangle_{R_1} |l\rangle_{C_1}  \Big)\\\nonumber
&&|l\rangle_{R_2}|k\rangle_{C_2}  |0\rangle_{B_1} |00\rangle_{B_2}|1\rangle_{B_3} +|g_3\rangle_{R_1C_1R_2C_2B_1B_2B_3} .
\end{eqnarray}
In this expression, \(|g_3\rangle_{R_1C_1R_2C_2B_1B_2B_3}\) (abbreviated as \(|g_3\rangle\)) aggregates all the components derived from the useless information items (originally labeled by \(|0\rangle_{B_3}\) in \(|\Phi_4\rangle\)) after the Hadamard transform. The factor \(\frac{1}{(\sqrt{2})^{3}}\) arises from the application of three independent Hadamard gates, each contributing a normalization factor of \(\frac{1}{\sqrt{2}}\).

To mark the newly generated useless information items, a 1-qubit auxiliary register \(B_4\) initialized in the state \(|0\rangle_{B_4}\) and a projection operator \(P_{B_1B_2B_3} = |0001\rangle_{B_1B_2B_3} \langle 0001|\) are introduced.
Subsequently, the following control operator is applied:
\begin{eqnarray*}
W^{(6)}_{B_1B_2B_3B_4} =P_{B_1B_2B_3} \otimes \sigma^{(x)}_{B_4} + (I_{B_1B_2B_3}-P_{B_1B_2B_3})\otimes I_{B_4}
\end{eqnarray*}
This control operator acts on the combined state \(|\Phi_5\rangle |0\rangle_{B_4}\), resulting in the state \(|\Phi_6\rangle\) which is expressed as:
\begin{eqnarray}
|\Phi_6\rangle &=& W^{(6)}_{B_1B_2B_3B_4}|\Phi_5\rangle |0\rangle_{B_4} \\\nonumber
&=&\frac{1}{(\sqrt{2})^{3}}\frac{1}{\sqrt{3}}\Big(\sum_{i=0}^{N-1} \sum_{j\neq l,k}  a_{ij}  |i\rangle_{R_1} |j\rangle_{C_1}   
+\sum_{i=0}^{N-1} a_{il}  |i\rangle_{R_1} |k\rangle_{C_1}   
+\sum_{i=0}^{N-1} a_{ik}  |i\rangle_{R_1} |l\rangle_{C_1}  \Big)\\\nonumber
&&|l\rangle_{R_2}|k\rangle_{C_2}  |0\rangle_{B_1} |00\rangle_{B_2}|1\rangle_{B_3}|1\rangle_{B_4} +|g_3\rangle_{R_1C_1R_2C_2B_1B_2B_3}  |0\rangle_{B_4}.
\end{eqnarray}
In this expression, the useful information items (the first term) are labeled with the state \(|1\rangle_{B_4}\) of auxiliary register \(B_4\), while the newly generated useless information items (aggregated in \(|g_3\rangle\)) remain labeled with \(|0\rangle_{B_4}\), achieving clear separation between the two types of information for subsequent operations.

By performing a measurement on the auxiliary register \(B_4\) with respect to the state \(|1\rangle_{B_4}\), we can eliminate the useless information items and retain only the useful ones. This measurement process yields the state \(|\Phi_7\rangle\), which is expressed as follows:
\begin{eqnarray}
|\Phi_7\rangle &=&
|\Psi_{out}\rangle |l\rangle_{R_2}|k\rangle_{C_2}|0001\rangle_{B_1B_2B_3} ,
\end{eqnarray}
where \(|\Psi_{out}\rangle\) denotes the output state corresponding to the useful information of the matrix, defined as:
\begin{eqnarray}
 |\Psi_{out}\rangle =\sum_{i=0}^{N-1} \sum_{j\neq l,k}  a_{ij}  |i\rangle_{R_1} |j\rangle_{C_1} 
+\sum_{i=0}^{N-1} a_{il}  |i\rangle_{R_1} |k\rangle_{C_1}+\sum_{i=0}^{N-1} a_{ik}  |i\rangle_{R_1} |l\rangle_{C_1}.
\end{eqnarray}

The measurement success probability of this algorithm is \(\frac{1}{24}\), and this probability is independent of both the elements \(a_{ij}\) of the matrix and the matrix dimensions (i.e., \(N\) and \(M\)).
Based on the preceding discussion, the overall computational complexity of this algorithm is determined by three key control operators: \(W^{(1)}_{R_2C_2B_1}\), \(W^{(2)}_{C_1R_2C_2B_2}\), and \(W^{(3)}_{C_1R_2C_2B_2}\).  Consequently, both the overall computational complexity and depth of this algorithm are \(O(m)\).

\textbf{Remark 2.} For the two types of elementary column transformations of a \(2^n \times 2^m\) matrix, they can also be realized using the matrix transposition and the two types of elementary row transformations proposed in our previous work \cite{LTLF}. 
To implement the transposition algorithm, one \(m\)-qubit ground state needs to be introduced, and \(m\) quantum gates are utilized, resulting in an algorithm complexity of \(O(m)\). The quantum algorithm complexity of the elementary row transformations is related to the row dimension of the matrix (after transposition, these "rows" correspond to the columns of the original matrix), with both the computational depth and complexity being \(O(m)\). Here, in comparison to the quantum algorithm that employs transposition and elementary row transformations to achieve elementary column transformations, the quantum algorithm that directly uses elementary column transformations of the matrix not only maintains the same computational complexity as the elementary row transformation algorithm but, more importantly, completely avoids the additional quantum resource consumption caused by the transposition operation.

\medskip

\vspace{0.2cm}

\section{Conclusion}

This study focuses on the implementation of quantum algorithms for matrix operations, specifically including quantum computing schemes for the Kronecker product (tensor product), Hadamard product (Schur product), and matrix column transformation operations (column addition and column interchange).
Among these, the quantum algorithm for the Kronecker product has a measurement success probability of 1, a computational depth of \(O(1)\), and a complexity of \(O(m)\); the quantum algorithm for the Hadamard product has a measurement success probability of \(\frac{G^2}{2^{(n+m)}}\), with both computational depth and complexity of \(O(\max\{n, m\})\); the quantum algorithm for column addition has a measurement success probability of \(\frac{G^2}{8}\) (and \(\frac{1}{24}\) for column swapping).

These quantum algorithms can be roughly divided into four key stages:
first, construct the initial quantum state required for the algorithm; second, design specific unitary transformation operations and introduce auxiliary qubits to realize the encoding and marking of target operation information; third, use Hadamard gate operations to prepare the encoded information items into quantum superposition states; finally, perform selective measurement on the auxiliary qubits to filter out invalid information items, thereby accurately retaining the useful information that meets the operation requirements in the output state.

\vspace{0.2cm}

\textbf{Acknowledgments}

This work is being supported by the following: the National Natural Science Foundation of China (NSFC) under Grants 11761073, 12075159, and 12171044; the Academician Innovation Platform of Hainan Province. We are deeply grateful to Prof. Junde Wu and Alexander I. Zenchuk for their active participation and invaluable contributions to this research.

\vspace{0.2cm}
\section{References}
\begingroup
\renewcommand{\section}[2]{}%

\endgroup 
\end{normalsize}


\begin{thebibliography}{99}
{\small
\bibitem{PW} Shor P W. {\it Algorithms for quantum computation: discrete logarithms and factoring}. Proceedings 35th Annual Symposium
 on Foundations of Computer Science. IEEE, 1994: 124-134.
\bibitem{PW1} Shor P W. {\it Polynomialtime algorithms for prime factorization and discrete logarithms on a quantum computer}. SIAM J. Comput. 1997:26, 5.
\bibitem{DD} Deutsch D. {\it Quantum theory, the Church¨CTuring principle and the universal quantum computer}. Proc. R. Soc. Lond. A, 1985, 400(1818): 97-117.
\bibitem{GLK} Grover L K. {\it A fast quantum mechanical algorithm for database search}. Proceedings of the twenty-eighth annual ACM symposium on Theory of computing, 1996: 212-219.
\bibitem{CD} Coppersmith D. {\it An approximate Fourier transform useful in quantum factoring}. IBM Research Report, 2002: RC 19642 .
\bibitem{WPF} Weinstein Y S, Pravia M A, Fortunato E M, Fortunato E M, Lloyd S, Cory D G. {\it Implementation of the quantum Fourier transform}.  Phys. Rev. Lett, 2001, 86(9): 1889.
\bibitem{NC} Nielsen M., Chuang I. {\it Quantum computation and quantum information}. Cambridge University Press, Cam
bridge, England, 2000.
\bibitem{WWL} Wang H F, Wu L A, Liu Y X, Nori F. {\it Measurement-based quantum phase estimation algorithm for finding eigenvalues of non-unitary matrices}. Phys. Rev. A. 2010, 82(6): 062303.

\bibitem{DGP} Duchamp G H E, Goodenough S, Penson K A. {\it Rational Hadamard products via Quantum Diagonal Operators}.  arXiv:2008, 0810.3641.
\bibitem{ZZR} Zhao L M, Zhao Z K, Rebentrost P, Fitzsimons J. {\it Compiling basic linear algebra subroutines for quantum computers}. Quantum Mach. Intell., 2021, 3(2): 21.
\bibitem{DMS} Datt M S. {\it Structure of the tensor product of two simple modules of quantum $GL_2$}.  arXiv: 2021, 2105.06615, .
\bibitem{THZ} Thies J, Hof M T, Zimmermann M, Efremov M. {\it Tensor product scheme for computing bound states of the quantum mechanical three-body problem}.  J. Comput. Sci, 2022, 64: 101859.
\bibitem{LL} W J Liu , Li Z X. {\it Secure and efficient two-party quantum scalar product protocol with application to privacy-preserving matrix multiplication}. IEEE Trans. Circuits Syst. I, 2023, 70(11): 4456-4469.
\bibitem{MAH} Mansuroglu R, Adil A, Hartmann M J, Holmes Z, Sornborger A T. {\it Quantum Tensor-Product Decomposition from Choi-State Tomography}. PRX Quantum, 2024, 5(3): 030306.
\bibitem{QZKW} Qi W T, Zenchuk A I, Kumar A, et al. {\it Quantum algorithms for matrix operations and linear systems of equations}. Commun Theor Phys, 2024, 76(3): 035103.
\bibitem{ZQKW} Zenchuk A I, Qi W T, Kumar A, et al. {\it Matrix manipulations via unitary transformations and ancilla-state measurements}. Quantum Inf Comput, 2024, 24(13-14): 1099-1109.
\bibitem{ZBQW} Zenchuk A I, Bochkin G A, Qi W T, et al. {\it Quantum algorithms for calculating determinant and inverse of matrix and solving linear algebraic systems}.  Quantum Inf Comput, 2025, 25(2): 195-215.

\bibitem{FZQW} Fel'dman E B, Zenchuk A I, Qi W T, et al. {\it Remarks on controlled measurement and quantum algorithm for calculating Hermitian conjugate}.  arXiv:2025, 2501.16028.

\bibitem{ZQW}
Zenchuk A I, Qi W T, Wu J D. {\it Arbitrary state creation via controlled measurement}. arXiv:2025, 2504.09462.


\bibitem{ZQW2}
Zenchuk A I, Qi W T, Wu J D. {\it Matrix encoding method in variational quantum singular value decomposition}. Quantum Inf Comput, 2025, 25(4): 356-368.

\bibitem{ZQW3}
 Fel'dman E B, Zenchuk A I, Qi W T, Wu J D. {\it Controlled measurement, Hermitian conjugation and normalization in matrix-manipulation algorithms}. arXiv:2025, 2504.00015.

\bibitem{LTLF} Liu Y H, Tao Y H, Lan Y K, Fei S M. {\it Quantum Algorithms for Matrix Operations Based on Unitary Transformations and Ancillary State Measurements}. arXiv:2025, 2501.15137.
\bibitem{HJ}    Horn R A, Johnson C R. {\it Topics in matrix analysis}. Cambridge university press, 1994.
\bibitem{LAJ} Laub A J. {\it Matrix analysis for scientists and engineers}. SIAM, 2004.

\bibitem{RC} Horn R A and Johnson C R.  {\it Matrix Analysis}. Cambridge university press, 2012.

\bibitem{SGPH} Styan G P H. {\it Hadamard products and multivariate statistical analysis}. Linear Algebra Appl, 1973, 6: 217-240.
\bibitem{KShV} Kitaev A Y, Shen A H, Vyalyi M N, {\it Classical and Quantum Computation}. Graduate
Studies in Mathematics, V.47, American Mathematical Society, Providence, Rhode Island
, 2002.}
\end{thebibliography}
\end{document}